%% file: main.tex
\documentclass[format=acmsmall, review=false, screen=true]{acmart}

\renewcommand\footnotetextcopyrightpermission[1]{} % removes footnote with conference information in first column
\usepackage{booktabs} % For formal tables

\usepackage[ruled]{algorithm2e} % For algorithms

\SetAlFnt{\small}
\SetAlCapFnt{\small}
\SetAlCapNameFnt{\small}
\SetAlCapHSkip{0pt}
\IncMargin{-\parindent}

\begin{document}
% Title portion. Note the short title for running heads
\title{A Survey of Distributed Denial of Service Attacks and Defenses}

\author{Rajat Tandon}
\affiliation{%
  \institution{University of Southern California, Information Sciences Institute}}

%\author{Genevieve Bartlett}
%\affiliation{%
%  \institution{University of Southern California, Information Sciences Institute}}
  
%\author{Jelena Mirkovic}
%\affiliation{%
%  \institution{University of Southern California, Information Sciences Institute}}

\begin{abstract}
A distributed denial-of-service (DDoS) attack is an attack wherein multiple compromised computer systems flood the bandwidth and/or resources of a target, such as a server, website or other network resource, and cause a denial of service for users of the targeted resource. The flood of incoming messages, connection requests or malformed packets to the target system forces it to slow down or even crash and shut down, thereby denying service to legitimate users or systems. This paper presents a literature review of DDoS attacks and the common defense mechanisms available. It also presents a literature review of the defenses for low-rate DDoS attacks that have not been handled effectively hitherto. 
\end{abstract}
\maketitle
\thispagestyle{empty}

\input{samplebody-journals}

\end{document}

%% file: samplebody-journals.tex
\section{Introduction}
Although different researches are fighting against DDoS attacks, even today, the deployment or development of effective methods and mechanisms of these researches cannot resist DDoS attacks. It is a hard problem to solve because it is multifaceted. There are multifarious ways of creating attacks that are extremely effective. Section~\ref{sec:classification} and ~\ref{sec:ddos} give an overview of the classification and types of the different DDoS attacks. Section~\ref{sec:defense} gives an overview of the commonly used defense mechanisms. I am interested in working on types of attacks that so far we have not been able to handle effectively. These are mainly low-rate attacks and their related works are mentioned in Section~\ref{sec:noteffective}.

\begin{figure}
\includegraphics[width=\columnwidth,scale=1]{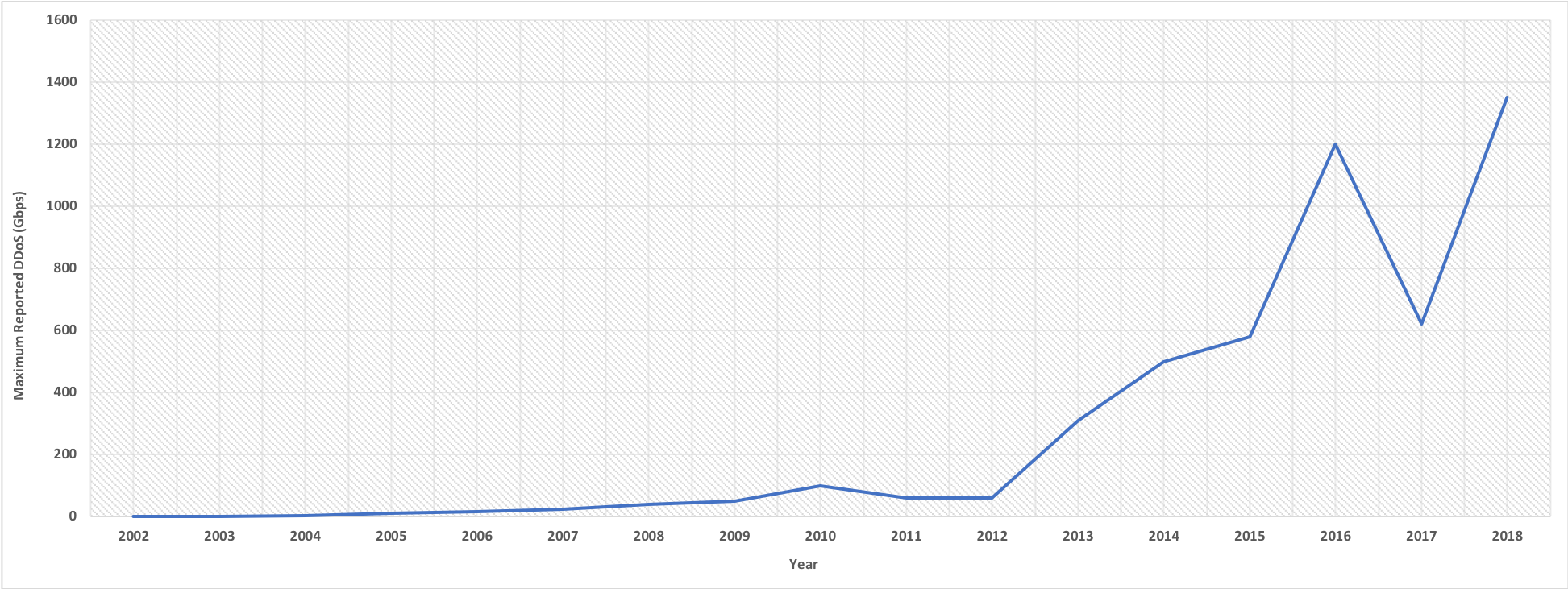}
\caption{Growth in DDoS Volume over the years \label{fig:ddos-growth}}
\end{figure}

DDoS attacks possess immense threat to the current Internet community, which comprises over 3.5 billion users \cite{IU}. Figure \ref{fig:ddos-growth} shows the year-wise growth statistics of DDoS attacks, collated from Arbor Networks: Worldwide Infrastructure Security Report \cite{Arbor}, Calyptix Security Blog \cite{Caplyptix}, DDoS attack on Dyn\cite{Dyn} in 2016 and the largest volumetric DDoS attack, ever recorded, on Github \cite{Github}. 
The 2016 Dyn cyberattack\cite{Dyn}, that took place on October 21, 2016 included multiple DDoS attacks targeting systems operated by Domain Name System (DNS) provider Dyn. It caused major Internet platforms and services to be unavailable to a large number of users spread across Europe and North America. And the magnitude of these attacks peaked to 1.2 Tbps.

The Github DDoS attack\cite{Github} generated a flood of internet traffic that peaked at 1.35Tbps, making it the largest on record. Interestingly, no botnets were involved in this attack. Instead, the hackers went with an amplification attack. They spoofed GitHub's IP address, and sent queries to several memcached servers that are typically used to speed up database-driven sites. The memcached systems then sent amplified responses from those requests to GitHub.

\begin{figure}
\includegraphics[width=\columnwidth,scale=1]{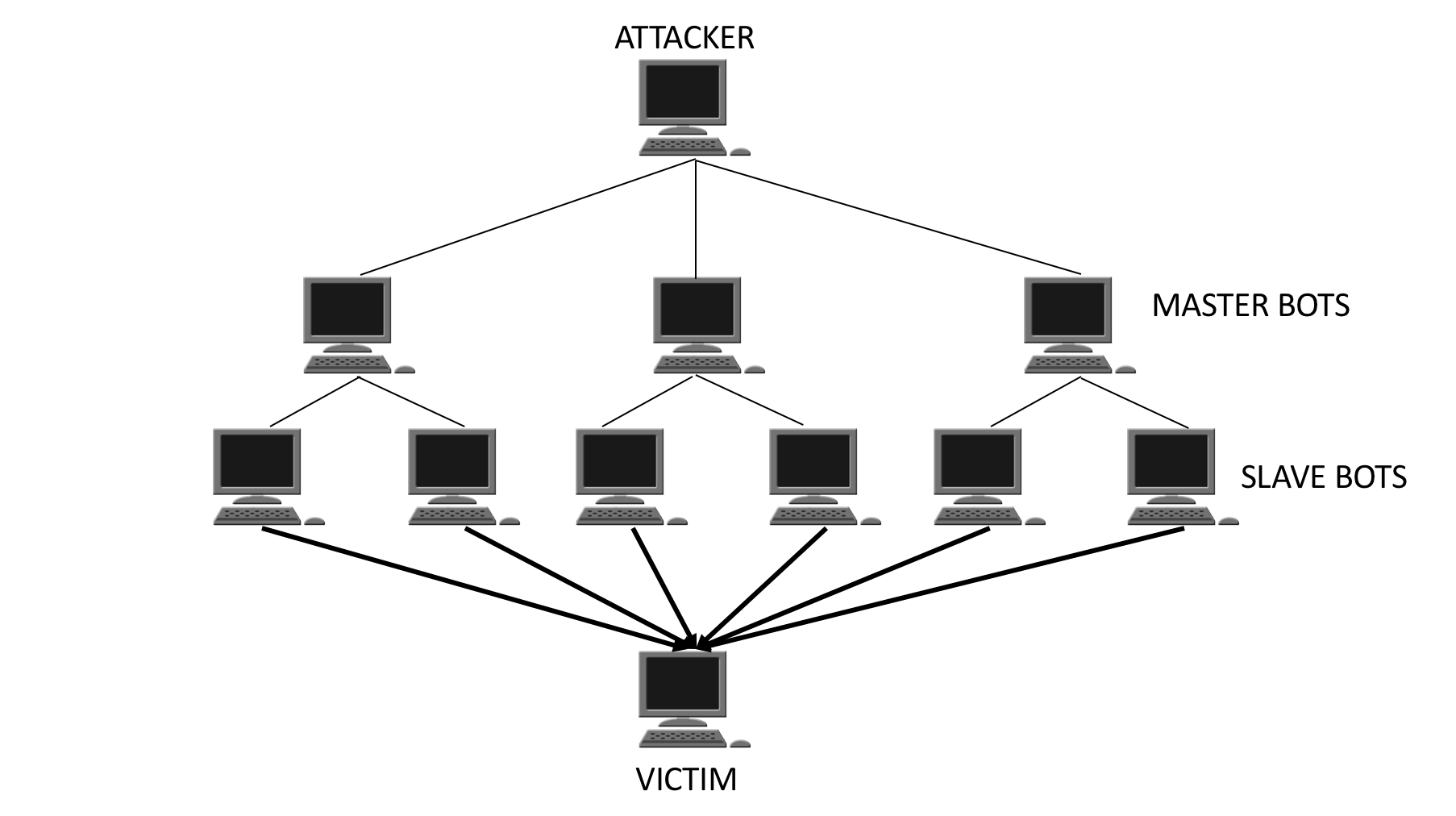}
\caption{DDoS Attack Architecture \label{fig:ddos-arch}}
\end{figure}

A computer or networked device under the control of an intruder is known as a bot. In general, in a DDoS attack, an attacker begins by exploiting a vulnerability in a computer system, making it the DDoS master bot. The attack master system identifies other vulnerable systems and gains control over them by either infecting the systems with malware or through bypassing the authentication controls. These systems are called DDoS slave bots. The attacker creates a command-and-control server to command the network of bots, referred to as a botnet.\cite{journals/ejwcn/LiuXGDZ09} Then, the attacker uses the traffic generated by the compromised devices to inundate the target so as to get its services down. Figure \ref{fig:ddos-arch} represents a typical DDoS architecture \cite{journals/cn/DouligerisM04}.

Also, the attackers aim that the DDoS attack packets do not reveal any obvious characteristic to segregate malicious and legitimate traffic. With minimal effort, the attackers aim to create the maximum-possible catastrophic impact.

The reasons or motivations behind DDoS attacks can be specific. But, in general, the major causes for DDoS attacks are as follows:

\begin{enumerate}
\item Competition: DDoS attacks can be intended to cripple company operations, damage reputation and devastate sales, which may directly benefit competitors.
\item Financial Gains: DDoS attacks can be used to achieve financial gains. Attackers can be hired, paid well or can even ask for ransom.
\item Politics: DDoS attacks have the potential to digitally silence opposition  parties. They can be used by political parties and terrorists. 
\item Revenge: Current employees, ex-employees, angry customers or anyone with a dispute may have a motive for a DDoS attack. 
\end{enumerate}

\section{Classification of DDoS Attacks}
\label{sec:classification}

There are a wide variety of possible DDoS attacks. And it is challenging to represent all of them with a single classification. Different literatures have classified DDoS attacks in different ways \cite{journals/ccr/MirkovicR04, conf/ISCApdcs/SpechtL04, journals/comsur/ZargarJT13,journals/nca/GuptaB17}. Arbor networks have classified DDoS attacks as volumetric attacks and application layer attacks. Volumetric attacks send a huge amount of traffic, or request packets, to a targeted network so as to overwhelm its bandwidth capabilities. Typically request sizes are in the 100's of Gbps. However, recent attacks have scaled to over 1Tbps. Application-layer attacks generally require a lot less packets and bandwidth to get a website down. They are silent and small, when compared to network-layer attacks, but extremely disruptive. 

\begin{figure}
\includegraphics[width=\columnwidth,scale=1]{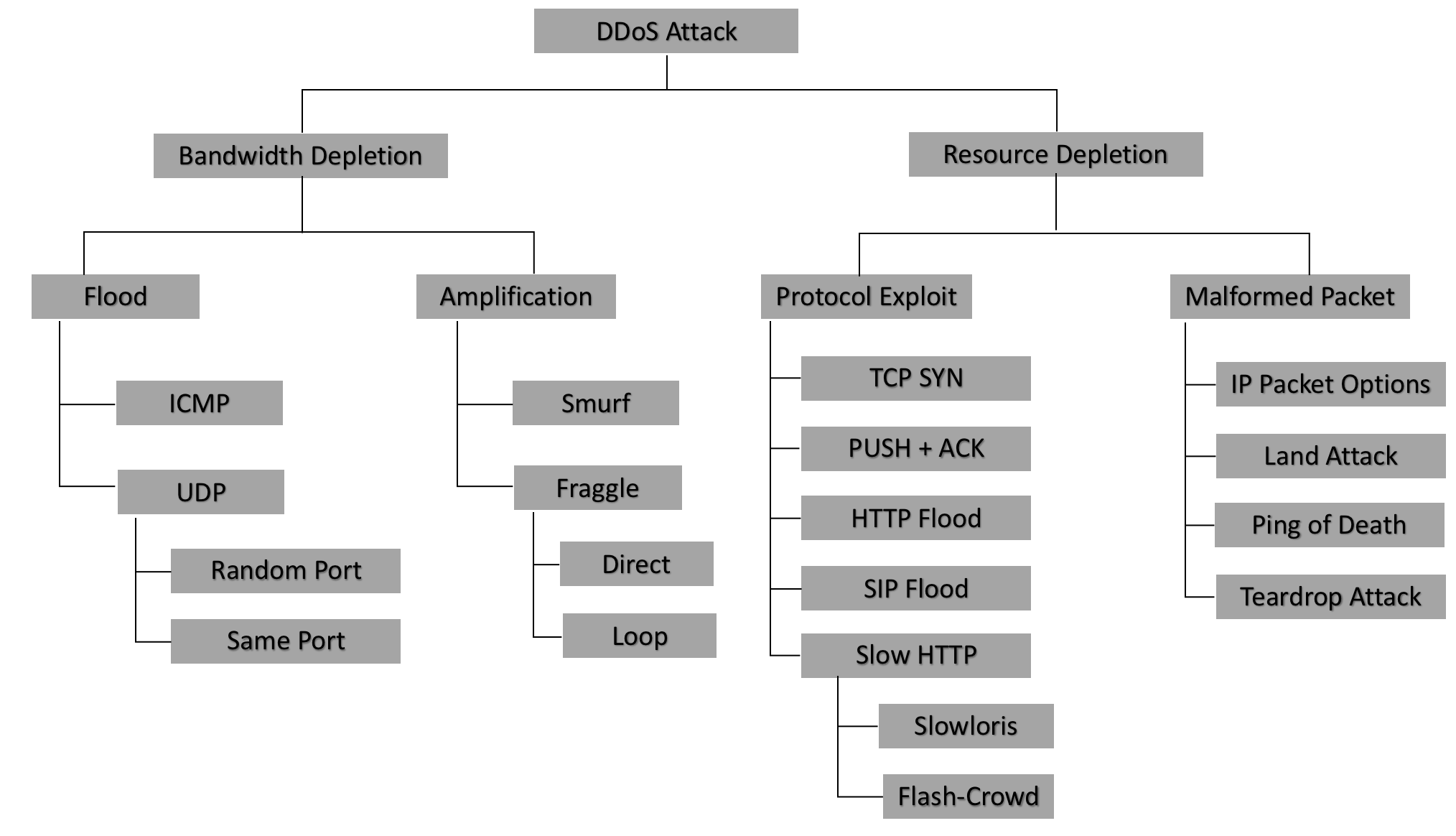}
\caption{Classification of DDoS Attacks \label{fig:ddos-classification}}
\end{figure}

Reference \cite{journals/ccr/MirkovicR04} presents the taxonomy of DDoS attacks and the different classifications include degree of automation, exploited weaknesses to deny service, source address validity, attack rate dynamics, possibility of characterization, persistence of agent set, victim type and impact on victim. Reference \cite{conf/ISCApdcs/SpechtL04} portrays a taxonomy of a few major DDoS attacks, shown in Figure \ref{fig:ddos-classification}. It includes two major classes of DDoS attacks: bandwidth depletion and resource depletion attacks. Bandwidth depletion attacks flood the victim
network with undesired traffic, preventing legitimate traffic from reaching the victim. And resource depletion attacks tie up the resources of a victim system making the victim unable to process legitimate requests.

\section{Different DDoS Attacks}
\label{sec:ddos}

The different DDoS attacks are enumerated below:
\subsection{Volumetric DDoS Attacks}

The volumetric DDoS attacks generate huge volumes of attack request or response traffic, usually measured in Gigabits per second (Gbps). The concept of a volumetric attack is simple, which is to send as much traffic as possible to a site to overwhelm its bandwidth. Some of the well known volumetric DDoS attacks are as follows:

\begin{enumerate}

\item DNS Amplification Attack: A DNS amplification attack is a DDoS attack in which the attacker exploits vulnerabilities in domain name system (DNS) servers to turn initially small queries into much larger payloads, which are used to bring down the victim’s servers. It is a reflection attack that manipulates publicly-accessible domain name systems, making them flood a target with large quantities of UDP packets. Using various amplification techniques, perpetrators can “inflate” the size of these UDP packets, making the attack so potent as to bring down even the most robust Internet infrastructure.

\item CharGEN Amplification Attack: A CharGEN amplification attack is usually carried out by sending small packets carrying a spoofed IP address of the target to the internet enabled devices running CharGEN. These spoofed requests to such devices are then used to send UDP floods as responses from these devices to the target. Internet-enabled printers and copiers have this protocol enabled by default and can be used to execute a CharGEN attack. This can be used to flood a target with UDP packets on port number 19. The server eventually exhausts its resources and goes down.

\item SNMP Amplification Attack: Simple Network Management Protocol(SNMP) is used for network management. SNMP amplification attack is carried out by sending small packets carrying spoofed IP address of the target to the internet enabled devices running SNMP. These spoofed requests to such devices are then used to send UDP floods as responses from these devices to the target. However, amplification effect in SNMP can be greater when compared with CHARGEN and DNS attacks. When the target tries to make sense of this flood of requests, it will end up exhausting its resources and will go down.

\item NTP Amplification Attack: Network Time Protocol(NTP) is used to synchronize the clocks of computers to some time reference. The NTP protocol is another publicly accessible network protocol. NTP amplification is essentially a reflection attack. Reflection attacks involve eliciting a response from a server to a spoofed IP address. The attacker sends a packet with a forged IP address, which belongs to the victim and the server replies to this address. The response is considerably larger than the request, amplifying the amount of traffic directed at the target server and ultimately leading to a degradation of service for legitimate requests.

\item SSDP Amplification Attack: The Simple Service Discovery Protocol(SSDP) is based on the Internet Protocol Suite for advertisement and discovery of network services and presence information. SSDP allows universal plug and play devices to send and receive information using UDP on port number 1900.  SSDP is attractive to DDoS attackers because of its open state that allows spoofing and amplification. The attacker conducts a scan looking for plug-and-play devices that can be utilized as amplification factors. As the attacker discovers networked devices, they create a list of all the devices that respond. The attacker creates a UDP packet with the spoofed IP address of the targeted victim. The attacker then uses a botnet to send a spoofed discovery packet to each plug-and-play device with a request for as much data as possible by setting certain flags, specifically ssdp:rootdevice or ssdp:all. As a result, each device will send a reply to the targeted victim with huge response.
The target then receives a large volume of traffic from all the devices and becomes overwhelmed and results in denial-of-service to legitimate traffic.

\item UDP Flood Attack: User Datagram Protocol(UDP) is a connectionless protocol that uses datagrams embedded in IP packets for communication without creating a session between two
devices. In a UDP Flood attack, a lot of UDP packets are sent to either random or specific ports on the victim system. In order to determine the requested application, the victim system processes the incoming data. In case of absence of the requested application on the requested port, the victim system sends a "Destination unreachable" message to the sender. Also, to conceal the identity, the attacker often spoofs the source IP address of the attacking packets. UDP flood attacks often deplete the bandwidth of network around the victim's system. Thereby, the systems close to the victim are also impacted due to the UDP flooding attack.

\end{enumerate}
\subsection{Protocol Attacks}

An internet protocol is basically a discrete set of rules to exchange information on the internet. These rules can be exploited by a bad actor. Some of the well known protocol attacks are as follows:

\begin{enumerate}
\item TCP SYN Flood Attack: The Transmission Control Protocol(TCP) SYN Flood attack is a DDoS attack that exploits the design of the three way TCP communication process between a client, host, and server. First, a client initiates a new session by sending a SYN packet to the server. The server responds with a SYNACK packet and then the client sends an ACK packet to the server before a connection is established. In a SYN flood attack, the attacker sends repeated SYN packets to every port on the targeted server, often using fake IP addresses. The server, unaware of the attack, receives multiple, apparently legitimate looking requests to establish communication. It responds to each of these with a SYN-ACK packet from each open port. Before the connection can time out, another SYN packet arrives. This leaves an increasingly large number of half-open connections. Ultimately, the server’s connection overflow tables get flooded and the service to legitimate clients are denied. Finally, the server may even malfunction or crash.

\item TCP SYN-ACK Flood Attack: The second step of the three-way TCP communication process is exploited by this DDoS attack. In a TCP SYN-ACK DDoS attack, attackers send spoofed SYN packets to a large numbers of servers and proxies on the Internet that generate large numbers of SYN-ACK packets in response to incoming SYN requests from the spoofed attackers. This SYN-ACK flood is not directed back to the botnet, but instead, is directed back to victim’s network and often exhausts the victim’s firewalls by forcing state-table lookups for every incoming SYN-ACK packet.  This denial of service attack can render stateful devices inoperable and can also consume excessive amounts of resources on routers, servers and IPS/IDS devices.

\item PUSH+ACK Flood Attack: When connecting with a server, the client can ask for confirmation that the information was received by setting the ACK flag, or it can force the server to process the information in the packet by setting the PUSH flag. Both requests require the server to do more work than with other types of requests. By flooding a server with spurious PUSH and ACK requests, an attacker can prevent the server from responding to valid traffic.

\item RST/FIN Flood Attack: After a successful three or four-way TCP-SYN session, RST or FIN packets are exchanged by servers to close the TCP-SYN session between a host and a client machine. In an RST or FIN Flood attack, a target server receives a large number of spoofed RST or FIN packets that do not belong to any session on the target server. This attack tries to exhaust a server’s resources as the server tries to process these invalid requests. The result is a server unavailable to process legitimate requests due to resources exhaustion.

\item IGMP Flood Attack: Internet Group Management Protocol(IGMP) is yet another connectionless protocol, used by IP hosts to report or leave their multicast group memberships for adjacent routers. An IGMP flood attack involves a large number of IGMP message reports being sent to a network or router, significantly slowing down and eventually preventing legitimate traffic from being transmitted across the target network.

\item ICMP Flood Attack/Ping Flood Attack: Internet Control Message Protocol (ICMP) is another connectionless protocol used for IP operations, diagnostics, and errors. Just as with a UDP flood, an ICMP flood does not rely on any specific vulnerability to achieve denial-of-service. An ICMP flood attack can involve an ICMP echo request. Once enough ICMP traffic is sent to a target server, it gets overwhelmed with requests, resulting in a denial-of-service. 

\item Multi-Vector Attack: Attackers can also combine more than one DDoS attack mechanisms to fire DDoS attacks keeping the engineers dealing with the DDoS attack confused. These attacks are the hard to discern and are capable of taking down some of the best-protected servers and networks. For example, HTTP Flood Attack can be combined with 
Recursive HTTP GET Flood Attack and fired together.

\item DNS Flood Attack:  In a Domain Name System(DNS) flood attack, the attacker targets one or more DNS servers belonging to a given zone, attempting to hamper resolution of resource records of that zone and its sub-zones. DNS servers help requesters find the servers they seek. A DNS zone is a distinct portion of the domain name space in the Domain Name System. The administrative responsibility is delegated to a single server cluster for each zone. In a DNS flood attack the offender tries to overbear DNS servers with apparently valid traffic, overwhelming server resources and impeding the servers' ability to direct legitimate requests to zone resources. The attacker sends multiple DNS requests to the victim’s DNS server directly or via a botnet. The DNS servers become flooded with requests,cannot process all of its incoming requests, eventually crashes.

\item Smurf Attack: The Smurf attack is a DDoS attack in which a large number of ICMP packets with the intended victim's spoofed source IP are broadcast to a computer network using an IP broadcast address. Most devices on a network will, by default, respond to this by sending a reply to the source IP address. If the number of machines on the network that receive and respond to these packets is very large, the victim's computer will be flooded with traffic. This can slow down the victim's computer to the point where it becomes impossible to work on.

\item Fraggle Attack: A Fraggle Attack is a DDoS attack that involves sending a large amount of spoofed UDP traffic to a router’s broadcast address within a network. It is very similar to a Smurf Attack.

\item SIP Flood Attack: The Session Initiation Protocol(SIP) flood attack is a DDoS attack that exploits the application layer protocol SIP, used in VOIP call setup. An attack can be made using different types of SIP request messages like SIP REQUEST, SIP INVITE or the SIP call control messages like SIP INFO, SIP NOTIFY, SIP RE-INVITE. The goal of this attack is to flood the proxy server or the SIP registration server and to consume all of its resources. Here, any army from the botnet sends thousands of messages to the SIP registrar server which is responsible to accept REGISTRAR requests as well as to keep records of the addresses and parameters of the user agents, unless it receives any server error message. As a result, the server overwhelms the legitimate users experience service outage and cannot reach the server.

\item IP Null Attack: Packets contain IPv4 headers which carry information about which transport protocol is being used. When attackers set the value of this field to zero, these packets can bypass security measures designed to scan TCP, IP, and ICMP. When the target server tries to put process these packets, it will eventually exhaust its resources and reboot.

\item HTTP Flood Attack: The real IP of the bots is used to avoid suspicion. The number of bots used to execute the attack is same as the source IP range for this attack. Since the IP addresses of the bots are not spoofed, there is no reason for defense mechanisms to flag these valid HTTP requests. A bot can be used to send a large number of GET, POST or other HTTP requests to execute an attack. Many bots can be collated in an HTTP DDoS attack to completely cripple the target server.

\item Single Session HTTP Flood: An attacker can exploit a loophole in HTTP 1.1 to send several requests from a single HTTP session. This allows attackers to send a large number of requests from a handful of sessions. In other words, attackers can bypass the limitations imposed by DDoS defense mechanisms on the number of sessions allowed. Single Session HTTP flood attacks target a server’s resources resulting in complete system shutdown or poor performance of the server.

\item Single Request HTTP Flood Attack: Single request HTTP Flood attack is based on the idea of sending several HTTP requests in a single HTTP session by masking these requests within one HTTP packet. When defense mechanisms evolved to block multiple incoming packets for a single session, attacks like Single Packet HTTP Flood were designed with workarounds to defeat these defenses. This evolution of an HTTP flood exploits another loophole in the HTTP technology. Several HTTP requests can be made by a single HTTP session by masking these requests within one HTTP packet. Keeping packet rates within the allowed limits, this technique allows an effective attack, exhausting a server’s resources. 

\item Recursive HTTP GET Flood Attack: For an attack to be highly successful, it must remain undetected for as long as possible. This is yet another kind of HTTP flood attack wherein the attacker requests multiple website pages, analyses replies, and then recursively requests each object of the website. This method can be used together with any kind of HTTP to make the attack as unnoticeable as possible. It is hard to detect as the recursive requests of the website objects resemble legitimate requests.

\item Random Recursive GET Flood Attack: This attack is a purpose built variation of Recursive GET attack. It is designed for forums, blogs and other websites that have pages in a sequence. Like Recursive GET it also appears to be going through pages. Since page names are in a sequence, to keep up appearance as a legitimate user, it uses random numbers from a valid page range to send a new GET request each time. Random Recursive GET also aims to deflate its target’s performance with a large number of GET requests leading to denial-of-service for actual legitimate users.

\item IP Packet Option Field Attack: This attack targets the optional fields of an IP packet and randomizes its value. For example, if in an IP packet, all qualities of service bits are set to 1, it causes the victim to apply some additional time to analyze the packet. Thus, it can inundate the processing ability of the victim if a flood of packets arrives with such deformation.

\item Ping of Death Attack: In a ping of death attack, an attacker intentionally forms a data packet that exceeds the maximum packet size which causes the victims to freeze or crash. This attack can be initiated by the attacker without requiring a botnet. It is an IP fragmentation attack that exploits the inherent size limitation that a packet can be transmitted in. 

\item Teardrop Attack: In this attack, the attacker  manipulates the offset value which in turn generates errors in fragmentation and reassembly of packets. The attacker sends fragmented packets with overlapping offset numbers. Thus, during the time of the packet re-assembly, invalid packets are created and crash the targeted server.

\item Session Attack: To bypass defenses, instead of using spoofed IP addresses, this attack uses the real IP address of the bots being used to carry out an attack. This attack is executed by creating a TCP-SYN session between a bot and the target server. This session is then stretched out until it times out by delaying the ACK packets. Session attacks try to exhaust a server's resources through these empty sessions. That, in turn, results in a complete system shutdown or unacceptable system performance.

\item UDP Fragmentation Flood Attack: It is another one of those cleverly masked DDoS attacks that are not easily detected. The activity generated by this attack resembles valid traffic and all of it is kept within limits. This version of the UDP Flood attack sends larger yet fragmented packets to exhaust more bandwidth by sending fewer fragmented UDP packets.When a target server tries to put these unrelated and forged fragmented UDP packets together, it will fail to do so. Eventually, all available resources are exhausted and the server goes down.

\item VoIP Flood Attack: This version of application specific UDP flood attack targets VoIP servers. An attacker sends a large number of spoofed VoIP request packets from a very large set of source IP. When a VoIP server is flooded with spoofed requests, it exhausts all available resources while trying to serve the valid and invalid requests.

This reboots the server or takes a toll on the server’s performance and exhausts the available bandwidth. VoIP floods can contain fixed or random source IP. Fixed source IP address attack is not easy to detect as it masks itself and looks no different from legitimate traffic.

\item ICMP Fragmentation Flood Attack: This version of ICMP Flood attack sends larger packets to exhaust more bandwidth by sending fewer fragmented ICMP packets. When the target server tries to put these forged fragmented ICMP packets with no correlation together, it will fail to do so. The server eventually exhausts its resources and reboots.
\end{enumerate}
\subsection{Low Rate DDoS Attacks}
\label{sec:low}
Detection of low rate DDoS attacks is extremely hard as they
possess the ability of concealing their traffic, because the traffic is
very much like the normal, legitimate traffic. They are able to elude the current
anomaly-based detection schemes. Some of the well known low rate DDoS attacks are as follows:

\begin{enumerate}
\item Flash-Crowd DDoS Attacks: A Flash Crowd Attack(FCA) is a DDoS attack that consumes the resources of a targeted service with legitimate looking requests generated by numerous bots. It is extremely hard to detect as bots may request legitimate content. An attacker may further employ several bots, each sending requests at a low rate.

\item TCP Shrew Attack: These are DDoS attacks which generate periodic, short bursts of high-volume traffic and create congestion. This forces legitimate TCP connections to drastically reduce their sending rate. Shrew attacks exploit, the deficiencies in the retransmission
time-out (RTO) mechanism of TCP flows. They throttle legitimate
TCP flows by periodically sending burst pulses with high peak
rate in a low frequency. As such, the TCP flows see congestion
on the attacked link every time it recovers from RTO. Indeed,
such a shrew attack may reduce the throughput of TCP applications
down to almost zero.

\item Hash Collisions DDoS Attacks: Most application servers create hash tables to index POST session parameters and are sometimes required to manage hash collisions when similar hash values are returned. Collision resolutions are
resource intensive, as they require an additional amount of CPU to
process the requests. In Hash Collision DDoS attacks, the
attacker sends a specially crafted POST message with a multitude
of parameters. These parameters are built in a way that causes hash
collisions on the server side. This slows down the response processing
dramatically. Hash Collisions DDoS attacks are very effective and can
be launched from a single attacker computer, slowly exhausting the
application server’s resources.

\item Costly Requests Attack: These are DDoS attacks wherein the attacker requests for content that requires costly database queries, which deplete bandwidth and connection pool between a front-end server and a database. The requests that involve long processing times, create a load on database processing and consume ample amount of server resources are selected by the attackers to launch the attack.

\item Slowloris Attack: A Slowloris DDoS attack works by opening multiple connections to the targeted web server. And it keeps them open as long as possible. This is achieved by continuously sending partial HTTP requests, none of which are ever completed. The attacked servers open more and connections open, waiting for each of the attack requests to be completed. As the requests are not completed, the targeted server’s maximum concurrent connection pool gets filled, and additional legitimate connection attempts are denied.

\end{enumerate}
\subsection{Vulnerability-Based DDoS Attacks}

Servers or protocols may have are unknown or unpatched  vulnerabilities. Attackers exploit these vulnerabilities to launch attacks. Examples of vulnerability-based DDoS attacks are:
\begin{enumerate}
\item Zero Day DDoS Attack: These are DDoS attacks that exploit new vulnerabilities. These ZERO Day DDoS vulnerabilities do not have patches or effective defensive mechanisms at the time of the attack.

\item LAND Attack: A LAND(Local Area Network Denial) attack is a DDoS attack that consists of sending a special poison spoofed packets to a computer, causing it to lock up.  Whenever a system receives this type of packets, it replies back to itself which in turn creates an infinite loop. As a result of this, the system crashes eventually.

\end{enumerate}

\section{DDoS Mitigation}
\label{sec:defense}

DDoS attacks pose immense threat to the resources of the victim and to the network bandwidth and infrastructure. Although for each attack, or a group of attacks, described in Section~\ref{sec:ddos}, there are specific defense mechanisms available in literature, some of the commonly used ways to mitigate DDoS attacks are as follows: 

\begin{enumerate}

\item Hop-count filtering:
A commonly used defense is hop-count filtering \cite{jin2003hop} that filters packets that contain spoofed IP addresses. In this approach, the authors have used the concept that it is not possible to alter the number of hops of an IP packet when it travels from a source to a destination. Therefore, the authors have used this count to determine the validity of a packet. In this method, time to live value is used to count the number of hops. This hop counts are stored in a mapping table against each source address. Upon receiving a packet, the number of hops required for this packet is calculated and matched against the mapping table. A packet is detected as spoofed packet if a mismatch is found in this comparison. If a flow of packets is identified as a flow of spoofed packets, the filter discards those packets as a prevention of an attack. The deployment of such a technique is easier as it requires implementation in the victim's system. But, it has a limitation in the process of hop count. As this method counts the number of hops based on time to live, the number of false positive is larger in this method. This is because the initial time to live value is usually different for different operating systems. Additionally, the attacker can forge valid hop counts in their packets which allows the packets to pass the filter. Finally, for a flood of malicious packets, the system cannot perform the calculation and comparison. Hence it becomes the victim of the DDoS attack.

\item Source Address Validity Enforcement protocol:
Source Address Validity Enforcement (SAVE) protocol \cite{li2002save} enforces the routers to send messages containing updated source information to each destination routers connected to a source. Then, each router updates its forwarding table with current information and uses it to filter the packets based on the methods of RPF. Thus, it overcomes the problem encountered in the RPF for the asymmetric and dynamic nature of the Internet. However, implementation of such protocol requires a change in the existing routing protocol which is a time-consuming process. Also, a partial deployment of the protocol does not guarantee full success in filtering spoofed IP addresses.

\item Rate Limiting:
Rate limiting is one of common ways to defend the DDoS attacks. Generally, if the results of a detection mechanism are found to be partially successful, that is, if it produces large false negatives or cannot identify a precise distinction between the legitimate and malicious traffic, it is reasonable to apply rate limiting rather than filtering.

\item Ingress/Egress filtering:
Ingress/Egress filtering is a commonly used technique to prevent traffic with spoofed IP addresses to enter into a protected network. Basically, ingress filtering filters the malicious traffic destined to a local network and egress filtering discards the malicious traffic leaving a local network. Ingress filtering defined in RFC 226768 allows the traffic to enter the network which matches with a predefined range of domain prefixes of the network. If an attacker uses spoofed IP address that do not match with the prefixes, it is discarded at the routers. Thus, this filtering technique ensures prevention from DDoS attacks where spoofed IP addresses are used. However, it is not a useful mechanism in the cases where the valid IP addresses of the botnets are used as a source IP addresses during the attack. Also, the success of these filtering depends on the knowledge of the range of expected IP addresses for a port which is not always achievable for the complicated topologies used in different networks. Moreover, if an attacker uses some spoofed IP addresses that fall into the valid address range, the filters in the routers cannot detect the malicious traffic in such scenarios. Also, for these filters, itunneling is required for mobile IP users so that they are not filtered by the routers using ingress/egress filtering. Moreover, as it does not ensure incentives to the Internet Service Providers, it is partially deployed in the networks. Additionally, many Internet Service Providers do not enforce this approach.

\item Load balancing:
The key function of a load balancer is to spread workloads across multiple servers to prevent overloading servers, optimize productivity, and maximize uptime. Load balancers also add resiliency by re-routing live traffic from one server to another if a server is under a DDoS attack. Therefore, load balancers help to eliminate single points of failure and reduce the attack surface. Load Balancing is an approach which tries to balance the loads of different systems so that no one system gets overloaded. The result of the load balancing helps to gain the optimal productivity and the maximum uptime. In order to ensure the maximum load balancing, a bandwidth increase is required on all critical connections. A good number of replicated servers and data centers are required to guarantee elimination of single point of failure.

\item Martian address filtering and source address validation:
Martian address filtering is defined in RFC 1812. It works for filtering spoofed IP addresses that are generated from a limited set of addresses. This filtering ensures that a router must not forward any packet which has an invalid source or destination IP address. These invalid IP addresses range from reserved or special IP addresses as well as the unallocated range of IP addresses. Additionally, it ensures that any packet with the destination IP address 255.255.255.255/32 must be discarded at the router.

Source address validation is also specified in RFC 1812 and implemented in the routers. In this filtering mechanism, the router compares the source address of a packet with the logical interface of the packet where it is received. If this interface does not match with the interface where the packet needs to be destined to reach the specified source address, the router discards the packet. Thus, it can filter packets with spoofed source IP addresses. However, the rate of the false positive may become high for the asymmetric routes of the Internet. This asymmetric nature does not guarantee the match of the interfaces upon receiving or return of a packet from a specific source address. Thus, this filtering may discard a large number of legitimate traffics. However, the essential issue here is that not all routers in the Internet implement these approaches.

\item Capability-based response:
This mechanism can help prevent flooding related DDoS attacks. The receiver cannot control how much traffic it would receive. There exists flow control and congestion control mechanisms, but a misbehaving sender does not care about it. Therefore, the techniques involved in capability-based DDoS response work to discern a solution to control such misbehaving sender. Stateless inter flow filter (SIFF)134 and traffic validation architecture (TVA)135 are two example methods of this type.

\item Route-based Packet Filtering:
Route-based Packet Filtering(RPF) filters packets with spoofed source IP addresses. This filtering technique enhances the scope of the ingress filtering by providing service to the core routers. It does provide filtering based on the route information of a packet in each link of a core router. It depends on the principle that each link of the core router accepts traffic from only a limited number of source addresses. Thus, an IP packet which has a different source address than this set of addresses is discarded by the core routers as it appears to be spoofed to the router. In order to implement this technique, it requires information of the Border Gateway Protocol routing topology. According to the simulation of Park and Lee,\cite{park2001effectiveness} a significant success of the technique will be achieved if 18\% of the autonomous systems implement this filtering technique. However, this number is found impractical in current Internet scenario. Additionally, in order to include the source addresses in the BGP message, it is required to modify the scheme used in BGP messaging. This increases the message size and processing time of the BGP messages. Moreover, if the routers do not maintain updated information, this technique can discard legitimate packets for unwanted root change. Also, as RPF filters packets are based on BGP messaging information, an attacker can deceive root information and filtering rules by stealing BGP sessions. Finally, the attacker can carefully choose IP addresses that do not resemble the spoofed IPs. This can make the method ineffective in protecting DDoS from spoofed IP. Reference \cite{duan2008controlling} extends the idea of Park and Lee\cite{park2001effectiveness} by designing a packet filtering mechanism which considers update messages of local BGP to filter out spoofed IP addresses. This method is easy to deploy on the current architecture which relies on the BGP routing protocol. It also reduces the rate of false positive and simplifies IP traceback process.

\item History-based filtering:
History-based filtering is a packet marking–based filtering technique where the history of the normal traffic is used to filter out the malicious traffic\cite{peng2003protection}. Here, the destination of an attack maintains a database of IP addresses. This database comprises IP addresses which are commonly found in the destination. Therefore, when a bandwidth attack is targeted to this system, the system only allows those IP addresses that appear in their database and discards all other IP packets. However, this filtering technique cannot successfully detect and discard the malicious flow if an attacker can simulate its attack traffic as a normal traffic.

\item Path identifier:
The path identifier method \cite{yaar2003pi} eliminates packets based on a path identifier that identifies the path of the attacker. Identifying the path is a deterministic approach where each packet is stamped with an identifier based on the path it has traveled. The packets that travel the same path contain the same identifier. Thus, if the victim can identify a packet traveling from the attacker, it can filter all the subsequent packets sent by the attacker. This approach works well when half of the routers get involved to mark the packets. Because it works with a small sized identification field, there remains the possibility that different paths can show the same path information. Therefore, it can give false-positives and false-negatives. Reference \cite{yaar2006stackpi} proposes an improved version of the path identifier approach called StackPi that improves path identifier's performance in terms of incremental deployment. Additionally the improved filtering mechanism is capable of identifying malicious flows based on just a single packet. As per the paper, this method can provide a reasonable amount of DDoS protection only if 20\% of the routers implement this marking scheme. However, this method does not consider the detection of spoofed IP address packets rather it marks and filters a malicious packet based on the deterministic packet marking mechanism. Moreover, a host or router looking for the path information through this method may need to have some supportive software as well as expense of processing making this method difficult to deploy, as mentioned in reference \cite{beverly2009understanding}.

\item PacketScore:
PacketScore \cite{kim2006packetscore} is a proactive filtering technique that uses Bayes theorem to compute conditional legitimate probability(CLP). The conditional legitimate probability is used to determine the likelihood of a legitimate packet based on the baseline profile value and the attribute value of a packet. The packet filtering works based on this CLP value and a dynamic threshold. As the filtering takes into account the statistical analysis, this method works well for new attack signatures as well as non spoofed attack traffics. However, this method requires a large amount of storage to deal with the increasing number of attack attributes. It introduces a DDoS defense scheme that supports automated online attack characterizations and accurate attack packet discarding based on statistical processing. The key idea is to prioritize a packet based on a score which estimates its legitimacy given the attribute values it carries. Once the score of a packet is computed, this scheme performs score-based selective packet discarding where the dropping threshold is dynamically adjusted based on the score distribution of recent incoming packets and the current level of system overload.

\item Secure overlay:
The idea behind secure overlay \cite{adkins2003towards} is to build up an overlay network on top of the IP network.
It uses three principles: (i) enabling end-hosts to communicate without revealing their IP address, (ii) giving end-hosts control to defend against Denial-of-Service (DoS) attacks at the overlay level, and (iii) making sure that the added functionality does not introduce vulnerabilities not present in the Internet. This overlay network is the entry point for the outside network to establish a communication to the protected network. It is assumed that the isolation can be achieved if a protected network hides its IP addresses or uses a distributed firewall. This firewall ensures that only trusted traffic from the nodes of the overlay network can get entry to the protected network. Although this mechanism ensures prevention from the DDoS attack, it is applicable to a private network.

\item Honeypots:
A honeypot \cite{weiler2002honeypots} attracts attackers to attack them. Therefore, the actual system remains protected. Also, honeypots  can be used to extract important information like records of attack activity, tools, and software used for the attack which can reveal information about an attacker. This information is further used to detect and mitigate a DDoS attack. But this technique also contains some drawbacks. For example, the static and passive nature of the honeypots do not guarantee complete concealing from the attacker.

\item Changing IP addresses:
Changing IP address technique \cite{geng2000defeating} allows the computer system changes its IP address to invalidate an old address which may be the potential target of the DDoS attacks. This approach works for an IP address based attack. However, this also incurs a lot of overheads such as updating different entry information. This method works well till the attacker does not get informed about the new IP address. This approach provides the technological futility of addressing the problem solely at the local level.

\item Congestion policing:
The goal of a bandwidth flooding attack is to congest the resource\cite{deng2010implementation}. The impact of this type of attacks can be reduced or eliminated by applying congestion policing mechanism. Re-feedback\cite{briscoe2005policing} and NetFence\cite{liu2010netfence} are two example mechanisms where congestion policing is applied to defend DDoS attacks. Re-feedback ensures metrics in data headers such as time to live and congestion notification will arrive at each relay carrying a truthful prediction of the remainder of their path. NetFence uses a secure congestion policing feedback, to enable robust congestion policing inside the network.

\item Applying security patches:
It is also required to update all security patches regularly to ensure that the system is not affected by bugs or worms. Also, in order to prevent IoT botnet’s generation, it is mandatory to change the default or generic passwords of the IoT devices.86 This is an important awareness from the users side that can fight against the massive IoT–based DDoS attacks. However, the irony is that most of the users of the IoT devices are not aware of the threats or even they do not know the name of the DDoS attacks. When an attacker compromises a device for the DDoS army, the user of the device does not notice any change in their performance or behavior. Thus, stealthily they help in the attack without even knowing the name of the attack that they are participating.

\end{enumerate}

\section{Defenses for Low Rate attacks}
\label{sec:noteffective}
Low-rate DDoS attacks are not handled effectively by the current defense mechanisms. These attacks have ability of concealing their traffic because they are very much like normal traffic. They have the capacity to elude the current anomaly-based detection schemes. Instead of exhausting the network bandwidth, these attacks deny service at the device level. They deplete the resources at the service or the operating system, making it difficult for
the device unable to process legitimate clients' traffic.
These attacks are challenging to detect at the network level since they generate requests at low rate. Examples of such attacks are Flash Crowd Attack, TCP Shrew Attack\cite{kuzmanovic2003low}, Slowloris Attack, Hash-Collision Attack, described in \ref{sec:low}. 

Instead of a generic approach, most defenses focus on just a single variant. There is no generic framework
to handle all low-rate DDoS attacks.

Some related work for the slow attacks is as follows:

\begin{enumerate}

\item Reference Architecture: Reference \cite{shtern2014towards} proposes a reference architecture
for defense against low-rate DDoS attacks, enabled by a Software
Defined Infrastructure(SDI). Here, both network
and computational resources are provisioned and modified
on demand to achieve the mitigation goals. The suspicious
traffic is detected and redirected to the Shark Tank system, which is
created on-demand. The Shark Tank system is a separate cluster which possesses full application capabilities designed
to monitor suspicious users. The key feature of the Shark Tank is to act
as a restricted area for the attacker so that they can be
placed under close surveillance. It does absorb
the damage from the DDoS attacks and allows the
system to learn from these attacks for future reference.

\item Collaborative Detection and Filtering: Reference \cite{chen2006collaborative} presents a defense by combining discrete Fourier transform (DFT) and a hypothesis test framework to cope with shrew attacks. By computing the autocorrelation sequence of sampled time series and converting them into frequency-domain spectrum using DFT, the authors find that the power spectrum density (PSD) of a traffic stream containing shrew attacks that has much higher energy in low-frequency band than that appears in the spectrum for legitimate TCP/UDP traffic streams.
Based on this distinction, we develop a distributed collaborative detection and filtering (CDF) scheme to detect and segregate the shrew attack flows from legitimate TCP/UDP traffic flows. In addition to software implementation, the scheme can be implemented by network processor or reconfigurable hardware. The DSP hardware pushes spectral analysis down to the lower packet-processing layer. If the packets are processed by hardware and malicious flows are filtered out timely, the router workload will not increase much in the presence of shrew attacks.

\item Randomization on TCP Retransmission Timeout (RTO): Reference \cite{yang2004defense} proposes randomization on TCP RTO as defense against such attacks. With RTO randomization, an attacker cannot predict the next TCP timeout and consequently cannot
inject the burst at the exact instant. The authors show that randomization can effectively mitigate the impact of DDoS attacks while maintaining fairness and friendliness to other connections.

\item HAWK: Reference \cite{kwok2005hawk} proposes a new stateful adaptive queue management technique called HAWK (Halting Anomaly with Weighted choKing) which works by judiciously identifying malicious shrew packet flows using a small flow table and dropping such packets decisively to halt the attack such that well-behaved TCP sessions can re-gain their bandwidth shares. The authors show that HAWK is highly agile.

\item Shrew Attack Protection: Reference \cite{chang2010taming}  presents a simple priority-tagging filtering
mechanism, called SAP (Shrew Attack Protection), that protects
well-behaved TCP flows against low-rate TCP-targeted
Shrew attacks. In this method, a router maintains a simple set of counters and keeps track of the drop rate for each potential victim. If the monitored drop rates are low, all packets are treated as normal and equally compete to be admitted to the output queue and only dropped based on the AQM (Active Queue Management) policy when the output queue is almost full. However, if the drop rate for a certain victim becomes higher than some dynamically determined threshold (called fair drop rate), the router treats packets for this victim as high-priority, and these high-priority packets are preferentially admitted to the output queue. SAP marks victim packets as high priority until their drop rate is below the fair drop rate. By preferentially dropping normal packets to protect high-priority packets, SAP can prevent low rate TCP-targeted Shrew attacks from causing a well-behaved TCP flow to lose multiple consecutive packets repeatedly. This technique protects well-behaved TCP flows away from near zero throughput when under an attack.

\item Treat as application-level attacks and discern anomalies: A way of detecting low rate DDoS attacks is to treat them as application-level attacks and look for anomalies
in the payload of the incoming service requests. This is the approach taken by DDoS defense providers
and requires costly deep-packet inspection. Reference \cite{jonker2016measuring} presents an investigation in the adoption of cloud-based DDoS Protection Service providers worldwide. The authors focus on nine leading providers according to namely Akamai, CenturyLink, CloudFlare,
DOSarrest, F5 Networks, Incapsula, Level 3, Neustar, and
Verisign. The investigation is done on the basis of long-term, active DNS measurements, which allows for a given domain name, to verify if traffic diversion towards a DDoS Protection Service is in place or not.

\item Feature Feature Score(FFSc): Reference \cite{hoque2016ffsc} introduces a statistical measure called Feature Feature Score
(FFSc) for multivariate data analysis to distinguish DDoS
attack traffic from normal traffic. The authors extract three features of
network traffic, entropy of source IPs, variation of source
IPs and packet rate to analyze the behavior of network traffic
for attack detection. They build profiles of these features during normal operation and detect attack traffic
as the traffic that deviates from these profiles.

\item New Information Metrics: Reference \cite{xiang2011low} proposes that the generalized entropy and information distance
metrics outperform the traditional Shannon entropy
and Kullback–Leibler distance metrics for the low-rate
DDoS attack detection in terms of early detection, lower
false positive rates, and stabilities. It proposes an effective IP traceback scheme based on an
information distance metric that can trace all attacks back
to their own local area networks (LANs) in a short time. The proposed generalized entropy metric can detect attacks several hops earlier than the traditional Shannon metric approach. The proposed information distance metric outperforms the popular Kullback-Leibler divergence approach as it can enlarge the adjudication distance and then obtain the optimal detection sensitivity.

\item Expectation of Packet Size(EPS): Reference \cite{zhou2017low} presents a
measurement, expectation of packet size, that is based on the distribution difference of the packet size to distinguish
low-rate DDoS attacks from legitimate traffic. The authors propose an
expectation of packet size measurement to distinguish
attack traffic from legitimate traffic. The proposed method
is independent of the attack pattern; therefore, it can avoid the inherent shortcomings of the signature-based metric. Moreover, this approach can achieve a large distance gap between the attack traffics and the legitimate traffics, which could further contribute to a low false-positive rate.

\item Traffic Cluster Entropy: Reference \cite{sachdeva2014traffic} proposes traffic cluster entropy as detection metric not only to detect DDoS attacks but also to distinguish DDoS attacks from Flash Events. The authors show that when flash events are triggered, source address entropy increases but the corresponding traffic
cluster entropy does not increase. However, when DDoS attack is launched, traffic cluster entropy also increases along with source address entropy.

\item E--LDAT: Reference \cite{bhuyan2016ldat} proposes E‐LDAT, a lightweight extended‐entropy metric‐based system for both DDoS flooding attack detection and IP traceback. It aims to identify DDoS attacks effectively by measuring the metric difference between legitimate traffic and attack traffic. IP traceback is performed using the metric values for an attack sample detected by the detection scheme. This approach uses a generalized entropy metric with packet intensity computation on the sampled network traffic with respect to time. The authors show that E‐LDAT works well for detecting four classes of DDoS flooding attacks, including constant rate, pulsing rate, increasing rate and subgroup attacks.

\item Machine Learning: Reference \cite{saied2016detection} proposes machine learning to detect and mitigate known and unknown DDoS attacks in real time environments. The authors chose an Artificial Neural Network (ANN) algorithm to detect DDoS attacks based on specific characteristic features (patterns) that separate DDoS attack traffic from genuine traffic.

\item Flow Correlation Coefficient: Reference \cite{yu2012discriminating} proposes a
discrimination algorithm using flow correlation coefficient as a similarity metric among suspicious flows. The authors present theoretical proofs for the feasibility of the proposed discrimination method. 

\item Multifractal Detrended Fluctuation Analysis(MF--DFA): Reference \cite{wu2016low} proposes the algorithm of multifractal detrended fluctuation analysis to explore the change in terms of
multifractal characteristics over a small scale of network traffic due to low rate DDoS attacks. Through wavelet analysis, the singularity and
bursty of network traffic under low rate DDoS attacks are estimated by using Holder exponent. The difference values, known as D--value, of Holder
exponent of network traffic between normal and under low rate DDoS attack situations are calculated. The D--value is used as the basis to
determine low rate DDoS attacks. A detection threshold is set based on the statistical results. The presence of low rate DDoS attacks can be confirmed
through comparing D--value with detection threshold. Experiments on detection performance have been performed in the test-bed
network and simulation platform. The extensive experimental results are congruent with the theoretical analysis.

\item FlowTrApp: Reference \cite{buragohain2016flowtrapp} proposes DDoS defense using FlowTrApp. Software Defined Network (SDN)  provides a central
control over the network which helps in getting the global view of the network. FlowTrApp which performs DDoS detection and mitigation using some bounds on two per flow based traffic parameters that is flow rate and flow duration of a flow. It attempts to detect attack traffic ranging from low rate to high rate and long lived to short lived attacks using an SDN engine consisting of sFlow based flow analytics engine sFlow-RT and an OpenFlow controller. 

\item K--Nearest
Neighbors traffic classification with correlation analysis (CKNN): Reference \cite{xiao2015detecting} presents a detection approach based on K-nearest
neighbors traffic classification with correlation analysis to detect DDoS attacks. The approach exploits correlation
information of training data to improve the classification accuracy and reduce the overhead caused
by the density of training data. Aiming at solving the huge cost, the authors also present a grid-based method named
r-polling method for reducing training data involved in the calculation.

\item  Matching Pursuit and Orthogonal Matching
Pursuit algorithms: Reference \cite{andrysiak2013ddos} proposes to use Matching Pursuit and Orthogonal Matching
Pursuit algorithms for DDoS detection. The major contribution of the paper is the proposition of 1D
KSVD algorithm as well as its tree based structure representation (clusters), that can be applied to detect low-rate DDoS attacks and network anomalies. The method maintains the features of malicious traffic in form of tree to improve the classification performance and to optimize the implementation in terms of time complexity.

\end{enumerate}

\section{Conclusions}

In this paper, we presented the classification and literature review of DDoS attacks. In spite of scrupulous research over the years to mitigate DDoS attacks, they still exist today, in fact, with larger intensities and impacts as there are different angles to consider. We also described some of the commonly used defense mechanisms. Also, we presented a literature review of the defenses for low-rate DDoS attacks that have not been handled effectively till today.

\begin{acks}

Special thanks to Dr. Jelena Mirkovic and Dr. Genevieve Bartlett for their valuable guidance and advice.

\end{acks}

% Bibliography
\bibliographystyle{ACM-Reference-Format}
\input{main.bbl}

%\bibliography{sample-bibliography}